\documentclass{ws-procs9x6}

\usepackage{epsfig}
\newcommand{\agt}{\,\rlap{\lower 3.5 pt \hbox{$\mathchar \sim$}} \raise 1pt
 \hbox {$>$}\,}
\newcommand{\alt}{\,\rlap{\lower 3.5 pt \hbox{$\mathchar \sim$}} \raise 1pt
 \hbox {$<$}\,}

\begin{document}

\title{Inclusive hadron electroproduction at HERA at NLO with and without
transverse-momentum constraint}

\author{B. A. Kniehl}

\address{II.~Institut f\"ur Theoretische Physik, Universit\"at Hamburg,\\
Luruper Chaussee 149, 22761 Hamburg, Germany\\
E-mail: kniehl@desy.de}

\maketitle

\abstracts{We study single-hadron inclusive electroproduction in $ep$
scattering at DESY HERA at next-to-leading order in the parton model of
quantum chromodynamics endowed with non-perturbative fragmentation functions.
Specifically, we consider charged-hadron production, with unspecified
transverse momentum $p_T$, in the Breit frame and $D^{*\pm}$ production as
a function of $p_T$, and perform comparisons with recent data from the H1
Collaboration.}

\section{Introduction}
\label{sec:one}

In the framework of the parton model of quantum chromodynamics (QCD), the
inclusive production of single hadrons is described by means of fragmentation
functions (FFs) $D_a^h(x,\mu)$.
At leading order (LO), the value of $D_a^h(x,\mu)$ corresponds to the
probability for the parton $a$ produced at short distance $1/\mu$ to form a
jet that includes the hadron $h$ carrying the fraction $x$ of the longitudinal
momentum of $a$.
Analogously, incoming hadrons and resolved photons are represented by
(non-perturbative) parton density functions (PDFs) $F_{a/h}(x,\mu)$.
Unfortunately, it is not yet possible to calculate the FFs from first
principles, in particular for hadrons with masses smaller than or comparable
to the asymptotic scale parameter $\Lambda$.
However, given their $x$ dependence at some energy scale $\mu$, the evolution
with $\mu$ may be computed perturbatively in QCD using the time-like 
Dokshitzer-Gribov-Lipatov-Altarelli-Parisi (DGLAP) equations.
Moreover, the factorisation theorem guarantees that the $D_a^h(x,\mu)$
functions are independent of the process in which they have been determined
and represent a universal property of $h$.
This entitles us to transfer information on how $a$ hadronises to $h$ in a
well-defined quantitative way from $e^+e^-$ annihilation, where the
measurements are usually most precise, to other kinds of experiments, such as
photo-, lepto-, and hadroproduction.
Recently, light-hadron FFs with complete quark flavour separation were
determined\cite{akk} through a global fit to $e^+e^-$ data from LEP, PEP, and
SLC thereby improving previous analyses.\cite{kkp,k}

In the following, we extend our previous report\cite{dis05} on the
electroproduction, through deep-inelastic scattering (DIS), of $\pi^0$ mesons
and charged hadrons with finite transverse momentum $p_T^\star$ in the
$\gamma^\star p$ c.m.\ frame at next-to-leading order (NLO)\cite{kkm} by
discussing charged hadrons with unspecified values of $p_T^\star$, including
$p_T^\star=0$, and $D^{*\pm}$ mesons with $p_T^\star>0$.

\section{Analytic Results}
\label{sec:two}

\begin{figure}[ht]
\begin{center}
\begin{tabular}{cc}
\parbox{0.45\textwidth}{\epsfig{file=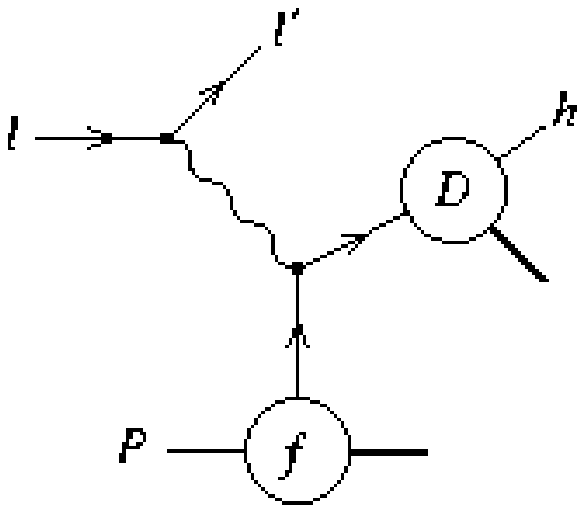,width=0.45\textwidth,%
bbllx=53pt,bblly=26pt,bburx=221pt,bbury=173pt,clip=}} & 
\parbox{0.45\textwidth}{%
\caption{(a) Parton-model representation of $l+p\to l^\prime+h+X$, with PDFs
($f$) and FFs ($D$), and Feynman diagrams for (b) virtual and (c) real NLO
corrections.
\label{fig:fig1}}}
\\
(a) & \\
\parbox{0.45\textwidth}{\epsfig{file=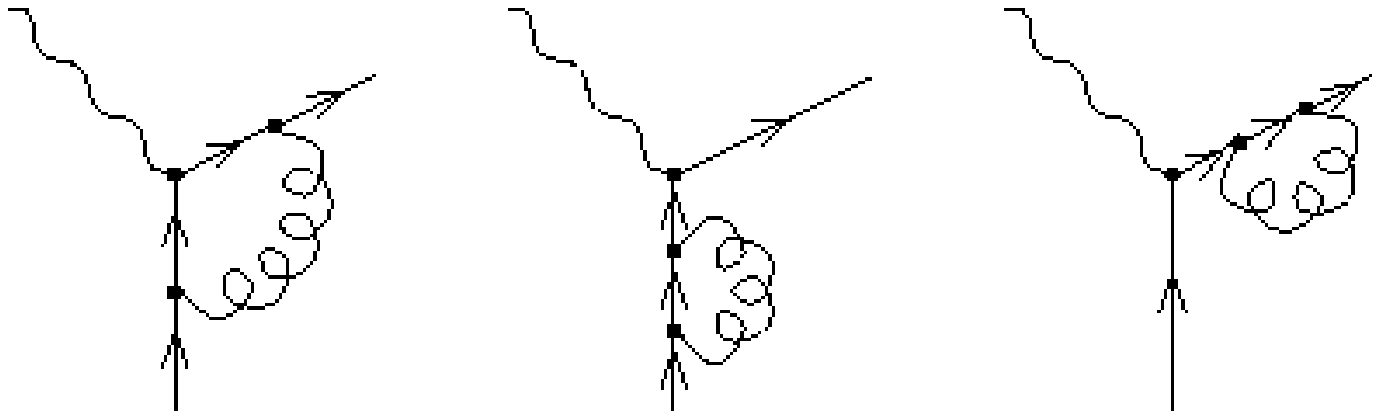,width=0.45\textwidth,%
bbllx=47pt,bblly=24pt,bburx=444pt,bbury=144pt,clip=}} &
\parbox{0.45\textwidth}{\epsfig{file=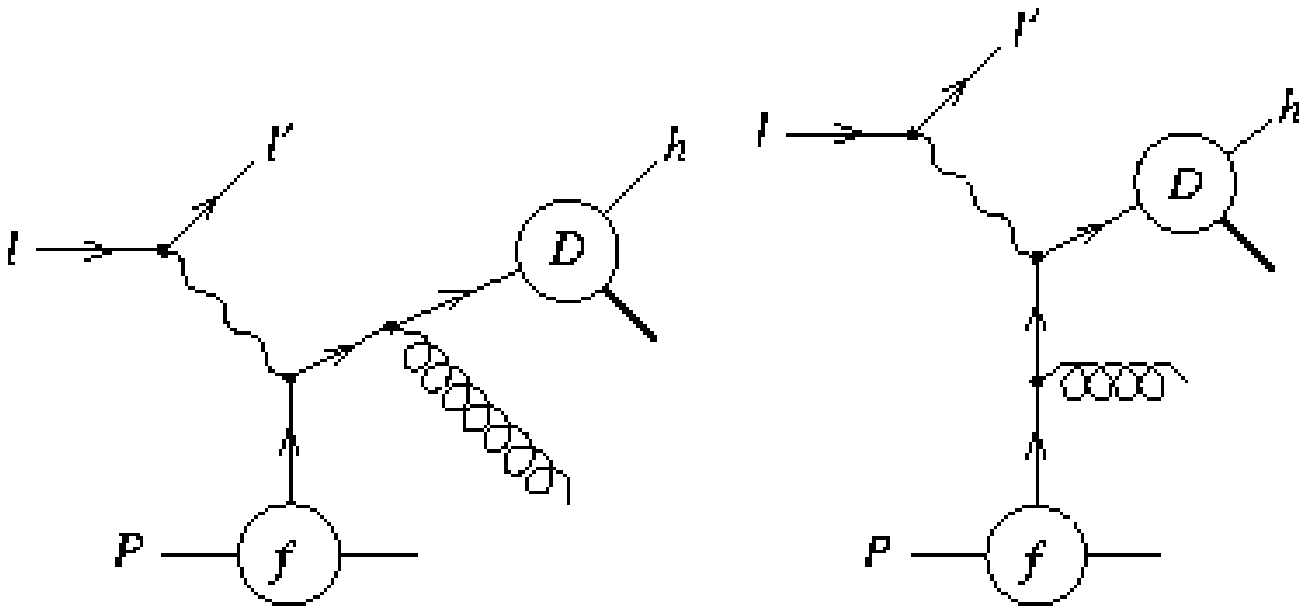,width=0.45\textwidth,%
bbllx=26pt,bblly=22pt,bburx=401pt,bbury=198pt,clip=}} \\
(b) & (c)
\end{tabular}
\end{center}
\end{figure}

At LO, inclusive hadron electroproduction proceeds through the Feynman diagram
shown in Fig.~\ref{fig:fig1}(a), so that $p_T^\star=0$.
At NLO,\cite{gra} virtual and real corrections, indicated in
Figs.~\ref{fig:fig1}(b) and (c), respectively, contribute.
In the latter case, $p_T^\star$ is integrated over.
The NLO cross section is conveniently evaluated with the {\tt FORTRAN} program 
{\tt CYCLOPS}.\cite{gra}

The NLO analysis for the case that $p_T^\star>0$ already at LO involves one
more external parton leg and may be found in Refs.~\refcite{kkm,aur}.

\section{Numerical Results}
\label{sec:three}

\subsection{Charged Hadrons in the Breit Frame}

H1\cite{h1} and ZEUS\cite{zeus} measured the normalised $Q$ distribution
$(1/\sigma_{\rm DIS})d\sigma/dQ$ of charged hadrons in bins of
$x_p=2p^{\rm Breit}/Q$, where $Q^2=-q^2$ is the virtuality of $\gamma^\star$
and $p^{\rm Breit}$ is the projection of the three-momentum of $h$ onto the
flight direction of $\gamma^\star$ in the Breit frame.
In this frame, $\gamma^\star$ is completely space-like, with four-momentum
$q^\mu=(0,0,0,-Q)$.
This frame provides a clear separation of current and remnant jets and is
especially appropriate for comparisons with inclusive hadron production by
$e^+e^-$ annihilation.

\begin{figure}[ht]
\begin{center}
\hspace*{2cm}
\begin{tabular}{cc}
\parbox{0.44\textwidth}{\hspace*{-2cm}
\epsfig{file=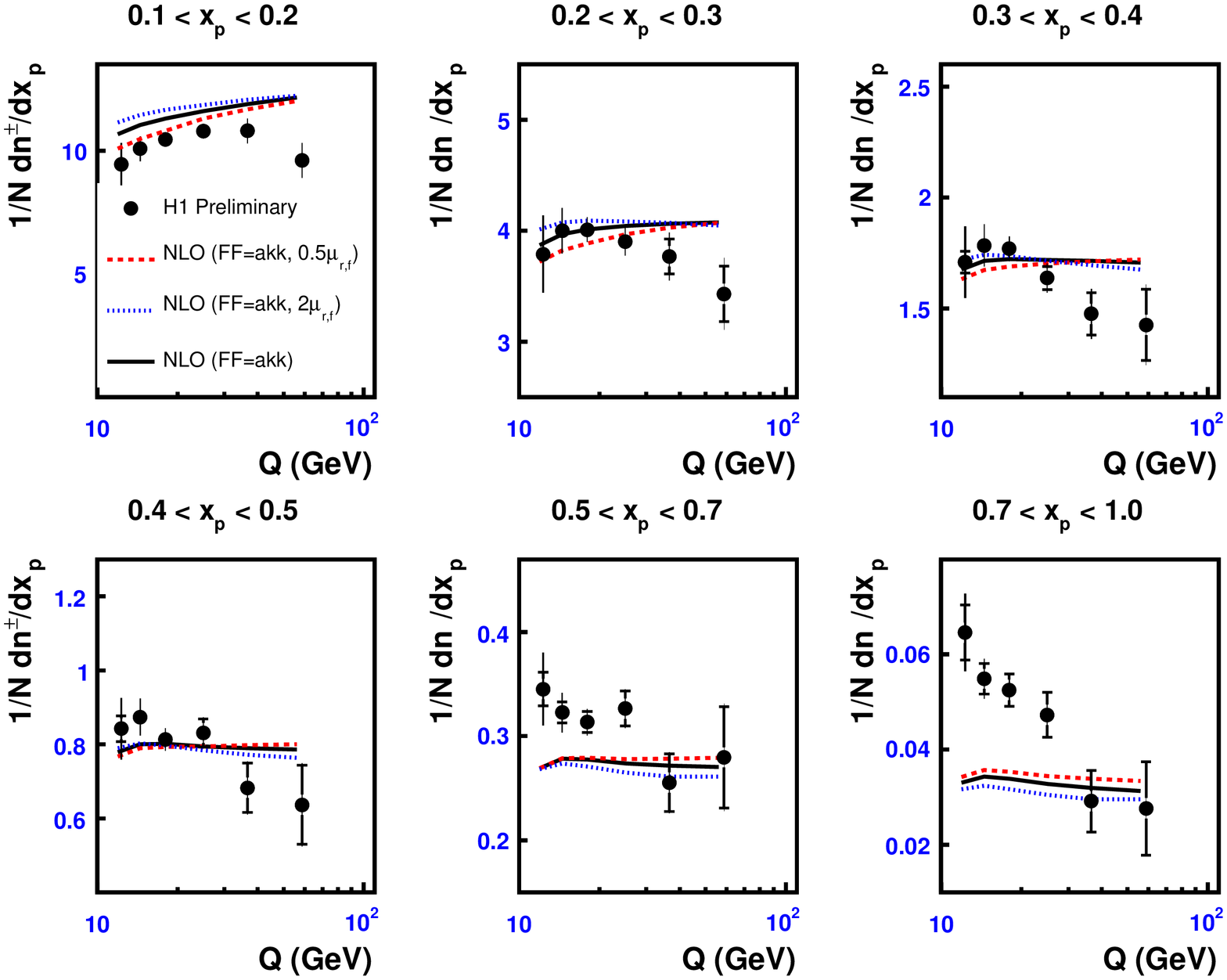,width=0.44\textwidth,%
bbllx=2,bblly=25,bburx=563,bbury=729,angle=270}} &
(a) \\
\parbox{0.44\textwidth}{\hspace*{-2cm}
\epsfig{file=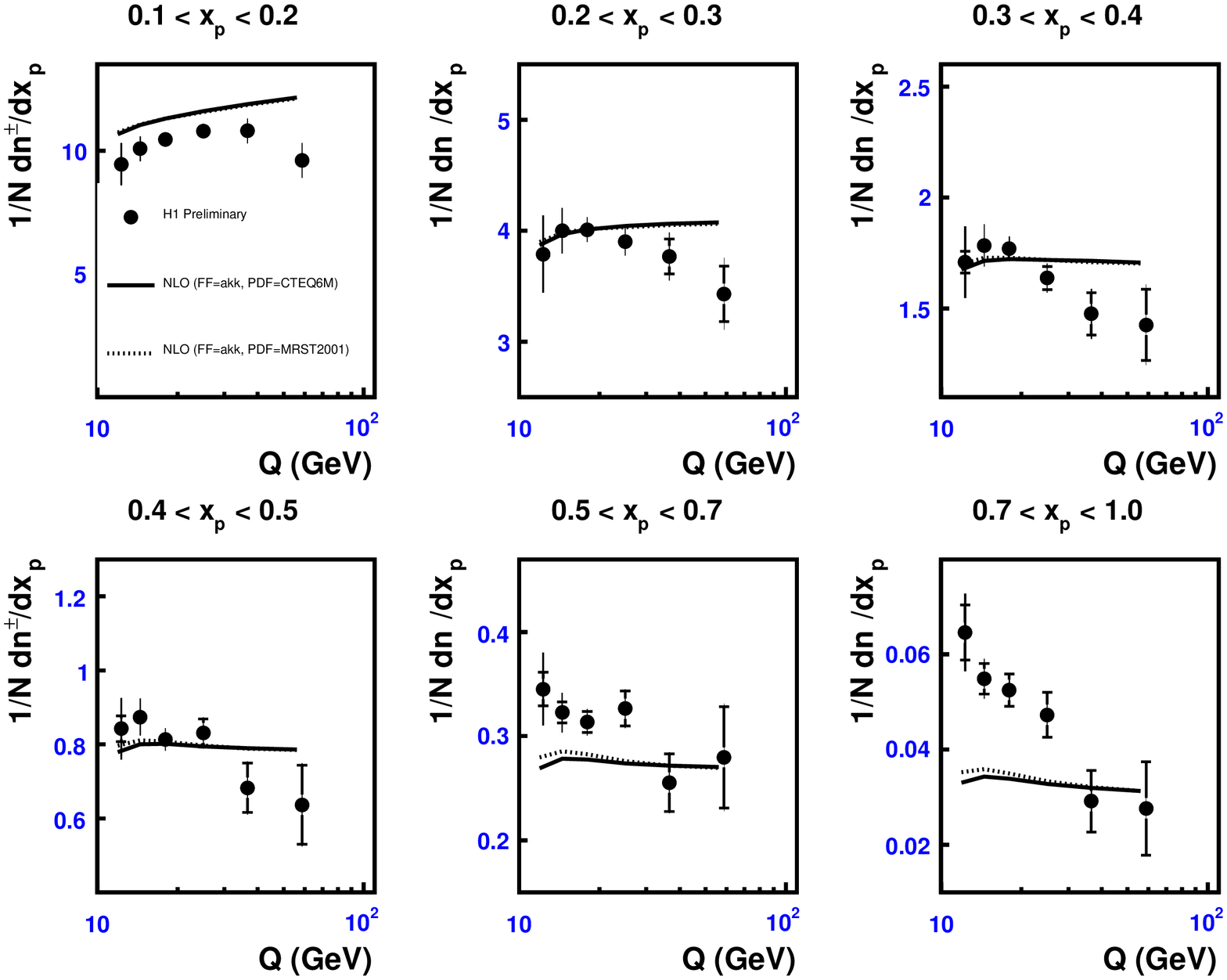,width=0.44\textwidth,%
bbllx=2,bblly=25,bburx=563,bbury=729,angle=270}} &
(b) \\
\parbox{0.44\textwidth}{\hspace*{-2cm}
\epsfig{file=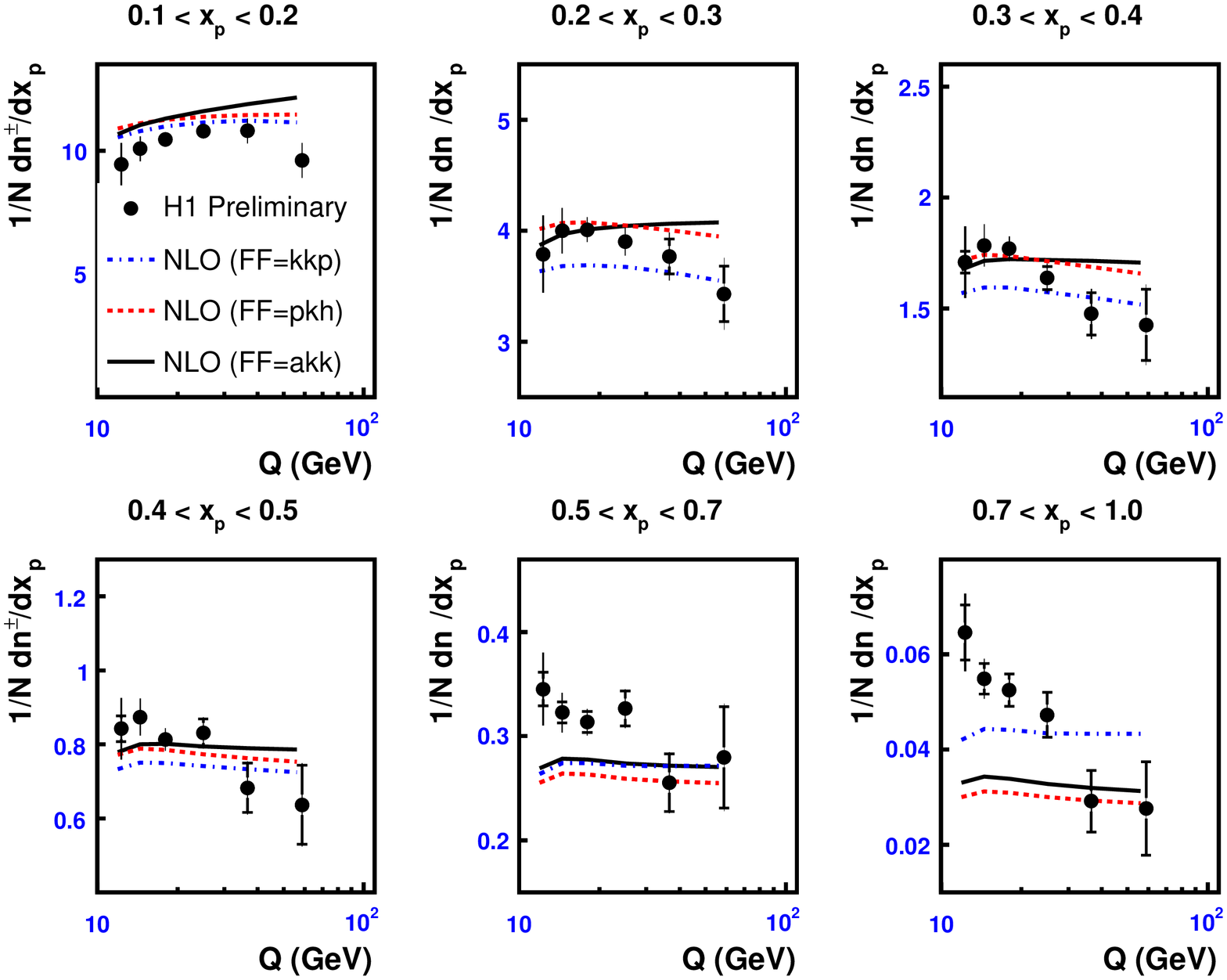,width=0.44\textwidth,%
bbllx=2,bblly=25,bburx=563,bbury=729,angle=270}} &
(c)
\end{tabular}
\end{center}
\caption{The normalised $Q$ distribution $(1/\sigma_{\rm DIS})d\sigma/dQ$ of
charged hadrons measured by H1\protect\cite{h1} in bins of $x_p$ is compared
with our NLO predictions estimating the theoretical uncertainties from the
freedom of choice of (a) unphysical scales, (b) PDFs, and (c) FFs.
\label{fig:fig2}}
\end{figure}

In Fig.~\ref{fig:fig2}(a), preliminary H1 data\cite{h1} are compared with NLO
predictions evaluated with CTEQ6.1M\cite{cteq} proton PDFs and AKK\cite{akk}
FFs; the renormalisation ($r$) and initial-state ($i$) and final-state ($f$)
factorisation scales are taken to be $\mu_r=\mu_i=\mu_r=\xi Q$, where $\xi$ is
varied between 1/2 and 2 about its default value 1 to estimate the
unphysical-scale uncertainty.
The PDF and FF uncertainties are assessed in Figs.~\ref{fig:fig2}(b) and (c)
by switching to the MRST2004\cite{mrst} PDFs and to the KKP\cite{kkp} and
K\cite{k} FFs, respectively.

\subsection{$D^{*\pm}$ Mesons}

Among other things, H1\cite{h1d} measured the $p_T^\star$ distribution
$d\sigma/dp_T^\star$ of $D^{*\pm}$ mesons in the DIS range $2<Q^2<100$~GeV$^2$
and $0.05<y<0.7$ with the acceptance cuts $p_T>1.5$~GeV and $|\eta|<1.5$ in
the laboratory frame, where $y$ is the relative lepton energy loss in the
proton rest frame and $\eta$ is the $D^{*\pm}$ pseudorapidity.

\begin{figure}[ht]
\begin{center}
\parbox{0.5\textwidth}{\epsfig{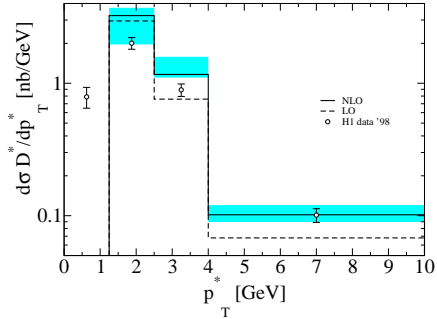}}
\end{center}
\caption{The $p_T^\star$ distribution $d\sigma/dp_T^\star$ (in nb/GeV) of
$D^{*\pm}$ mesons measured by H1\protect\cite{h1d} is compared with our LO and
NLO predictions.
\label{fig:fig3}}
\end{figure}

In Fig.~\ref{fig:fig3}, H1 data\cite{h1d} are compared with LO and NLO
predictions evaluated with CTEQ6\cite{cteq} proton PDFs, BKK\cite{bkk} FFs,
and $\mu_r^2=\mu_i^2=\mu_r^2=\xi\left[Q^2+\left(p_T^\star\right)^2\right]/2$
for $\xi=1$.
The theoretical uncertainty at NLO is estimated by varying $\xi$ between 1/2
and 2 about its default value 1.

\section{Conclusions}
\label{sec:four}

We compared H1 data on the electroproduction of charged hadrons\cite{h1} in
the Breit frame and of $D^{*\pm}$ mesons\cite{h1d} with $p_T^\star>0$ with
up-to-date NLO predictions.
In the first case, we found reasonable agreement, except for the region of
$Q\alt30$~GeV and $x_p\agt0.5$, where the FFs are generally less well
constrained by $e^+e^-$ data.
In the second case, we found good agreement for $p_T^\star\agt1.25$~GeV.
This nicely supports the scaling violations in the FFs encoded via the DGLAP
evolution as well as their universality predicted by the factorisation
theorem. 

\section*{Acknowledgements}

The author thanks G. Kramer, M. Maniatis, and C. E. Sandoval Usme for their
collaboration in the work presented here.
This work was supported in part by BMBF Grant No.\ 05~HT4GUA/4.

\end{document}